\documentclass[12pt]{iopart}
\usepackage{iopams}
\usepackage{amstext}
\usepackage{amssymb} 
\usepackage{latexsym}
\usepackage[english]{babel}
\usepackage[dvips]{graphicx}
\usepackage{color}

\newcommand{\beq}{\begin{equation}}
\newcommand{\eeq}{\end{equation}}
\newcommand{\ket}[1]{\left\vert#1\right\rangle}
\newcommand{\bra}[1]{\left\langle#1\right\vert}
\newcommand{\Ham}{\mathcal H}
\newcommand{\sandwich}[3]{\langle#1|#2|#3\rangle}

\begin{document}

\title[Many-body localization and thermalization in the FPDF of observables]
{Many-body localization and thermalization \\ in the full probability distribution function of observables}

\author{Elena Canovi$^{1}$, Davide Rossini$^2$, Rosario Fazio$^2$, Giuseppe E. Santoro$^{3,4,5}$ and Alessandro Silva$^{4}$}

\address{$^1$ \, Institut f\"ur Theoretische Physik III, Pfaffenwaldring 57,  D-70569, Stuttgart, Germany}
 
\address{$^2$ \, NEST, Scuola Normale Superiore, and Istituto Nanoscienze\,-\,CNR,
  Piazza dei Cavalieri 7, I-56126 Pisa, Italy}
  
\address{$^3$ \, SISSA, Via Bonomea  265, I-34136 Trieste, Italy}

\address{$^4$ \, International Centre for Theoretical Physics (ICTP), P.O. Box 586, I-34014 Trieste, Italy}
  
\address{$^{5}$\, CNR-IOM Demoscritos National Simulation Center, Via Bonomea 265, I-34136 Trieste, Italy}
  

\date{\today}

\begin{abstract}
  We investigate the relation between thermalization following a quantum quench and many-body localization
  in quasiparticle space in terms of the long-time full distribution function of physical observables. 
  In particular, expanding on our recent work [E. Canovi {\em et al.}, Phys. Rev. B {\bf 83}, 094431 (2011)],
  we focus on the long-time behavior of an integrable XXZ chain subject to an integrability-breaking 
  perturbation. After a characterization of the breaking of integrability and the associated 
  localization/delocalization transition using the level spacing statistics and the properties 
  of the eigenstates, we study the effect of integrability-breaking on the asymptotic state 
  after a quantum quench of the anisotropy parameter, looking at the behavior 
  of the full probability distribution of the transverse and longitudinal magnetization of a subsystem.
  We compare the resulting distributions with those obtained in equilibrium at an effective temperature
  set by the initial energy. We find that, while the long time distribution functions appear to always agree 
  {\it qualitatively} with the equilibrium ones, {\it quantitative} agreement is obtained 
  only when integrability is fully broken and the relevant eigenstates are diffusive in quasi-particle space. 
\end{abstract}



\section{Introduction}

The physics of thermalization in isolated quantum systems has a long and debated history.
Recent groundbreaking experiments on the non-equilibrium dynamics of low-dimensional 
condensates~\cite{Greiner2002,Sadler2006,Kinoshita2006} triggered a great deal of attention
on this topic, which, up to them, was mostly addressed as an academic question,
in connection with the notion of quantum 
chaos~\cite{deutsch91, srednicki94, Zelevinsky_PR96, Flambaum_PRE97a, Jacquod_PRL97, Benenti01}.
The demonstration of the lack of thermalization in two colliding bosonic clouds
confined in a cigar-shaped potential~\cite{Kinoshita2006}, and the attribution 
of this observation to quantum integrability, generated a lot of theoretical activity devoted 
to the study of the connections between integrability, ergodicity and thermalization 
in strongly correlated quantum systems~\cite{Polkovnikov2011,Lamacraft2011}.
The main focus efforts has been on the characterization of thermalization 
resulting from the simplest possible non-equilibrium protocol: an abrupt change in time 
of one Hamiltonian control parameter, that is a {\it quantum quench}. 

At long times after the quench, the lack of thermalization in an integrable system 
can be seen as a consequence of the sensitivity to the specifics of the initial state, 
that are encoded in the values of the constants of motion along the whole time evolution. 
Following the prescriptions of Jaynes~\cite{Jaynes1957}, this qualitative understanding 
was made rigorous by the proposal of describing the steady state after a quench 
by means of a generalized Gibbs ensemble, which keeps track of the initial value of all 
the non trivial constants of motion~\cite{Rigol_PRL07}. 
The conditions of applicability and drawbacks of this approach have been extensively 
tested (see Ref.~\cite{Polkovnikov2011} and references therein).
If the system is in turn far enough from the integrable limit, thermalization is expected in general to occur. 
This expectation is based on the eigenstate thermalization hypothesis, stating that 
expectation values of few-body observables in a given eigenstate equal thermal averages 
with the corresponding mean energy~\cite{deutsch91, srednicki94}, and it has been tested 
by means of several numerical techniques (see Refs.~\cite{Rigol_NAT08, Polkovnikov2011} and references therein).
The issue is still under debate~\cite{Roux_PRA09, Biroli_PRL10, Brandino_PRB12}.

A natural scenario to describe the effects of integrability and its breaking on thermalization is that of 
{\it many-body localization}. Building on a series of seminal papers in disordered 
electron systems~\cite{Altshuler_PRL97, Basko_AP06, Gornyi_PRL05}, the interplay 
between integrability breaking and many-body localization has been recently studied 
in the context of thermalization~\cite{Pal_PRB10, Canovi2011, Carleo2012,Khatami_PRE12}, 
In analogy to a construction originally conceived for disordered 
electron systems, the quasi-particle space can be thought of as a multidimensional lattice 
where each point is identified by the occupations of the various quasi-particle modes.
As long as states are localized in quasi-particle space, the system behaves as integrable: 
any initial condition spreads into few sites maintaining strong memory of the initial state.
Thermalization will not occur. 
At the same time, qualitative behavior of local and non-local operators in the quasi-particle 
is naturally going to be different: 
locality in quasi-particle space implies sensitivity to the localization/delocalization of states, 
while non-local operators display always an effective asymptotic thermal behavior. 
Once a strong enough integrability-breaking perturbation hybridizing the various states 
is applied, the consequent delocalization in quasi-particle space will lead to thermalization. 
An initial state is allowed to diffuse into all states in a micro-canonical energy shell generating 
a cascade of all possible lower energy excitations.

While this scenario appears physically sound, it does not give information on what is 
the degree of sensitivity of the various physical quantities of a many-body quantum system 
to integrability breaking and thermalization.
This question is particularly important in view of the fact that recent studies on the dynamics 
of quantum field theories lead to the proposal of a dynamics of thermalization 
comprising two stages~\cite{Berges2004}: first the system decays to a so-called {\it pre-thermalized} 
state, where the expectation value of certain macroscopic observables is to a good 
approximation ``thermal'', while the distribution function of the elementary degrees of freedom 
is not~\cite{Berges2004, Moeckel_PRL08, Moeckel_NJP10}. 
At a second stage, when the energy is efficiently redistributed by scattering processes,
real thermalization eventually occurs. 
Pre-thermalization has been shown to occur theoretically for quantum quenches in a variety of 
systems~\cite{Moeckel_PRL08, Moeckel_NJP10, Eckstein2009, Kollar2011,Barnet2011, Mathey2010, Marino2012}).
Moreover, the study of pre-thermalization in weakly perturbed integrable systems shed light 
on the nature of the pre-thermalized state, which is nothing but a close relative of the 
non-thermal steady state attained asymptotically by its integrable counterpart~\cite{Kollar2011}. 
Signatures of pre-thermalization have been observed in split one dimensional condensates~\cite{Gring2011},
which have been shown to be characterized by an intermediate, pre-thermalized stationary state.
The latter has been investigated by studying the full probability distribution 
of the interference contrast~\cite{Gring2011, Kitagawa2011} which turns out to have a closely 
{\it thermal} behavior, even though the distribution of quasi-particles is {\it non-thermal}.

The purpose of this paper is to provide a detailed characterization of the effect 
of integrability breaking on the asymptotic state after a quantum quench, not only by studying 
the average expectation values of certain selected observables, but also focusing on their 
{\it full probability distribution function} (FPDF), following the suggestions
that were recently put forward in Ref.~\cite{Kitagawa2011}.
We stress that the latter quantity is experimentally accessible by studying shot-to-shot 
variations of a physically measurable observable, as it has been performed 
for the FPDF of matter-wave interference in a coherently split Bose gas~\cite{Gring2011}.
In the first part we introduce our model and discuss the long time limit
of two kind of observables, thus summarizing the results contained in Ref.~\cite{Canovi2011}; 
in the second part we take a considerable step forward, by extending the discussion to quantities 
which are closer to actual experiments~\cite{Gring2011}, 
and clarifying more precisely to what extent thermalization takes place. 
We will show that, in the presence of many-body localization, the entire probability distribution function 
describes a canonical distribution of the degrees of freedom.

In the specific, we consider a quantum XXZ spin-$1/2$ chain undergoing a sudden quench of the anisotropy,
in the presence of an integrability breaking term in the form 
of a random transverse field. As the strength of the integrability breaking term is cranked up, 
the many-body level statistics has a well defined transition from Poisson (Integrable) 
to Wigner-Dyson (non-Integrable), closely associated to the localized/diffusive character 
of eigenstates in quasi-particle space~\cite{Canovi2011}. 
Focusing on the asymptotic state attained after a quench from the antiferromagnetic 
to the critical phase, we compute the FPDF of both transverse and longitudinal magnetization 
densities in a given spatial interval by means of exact diagonalization techniques. 
We compare the results with the FPDF of the same observables obtained in a canonical ensemble, 
with an effective temperature fixed by the initial energy. 
We show that, while for both weak and strong integrability breaking the FPDFs attained 
after a quench agree {\it qualitatively} with those obtained in the corresponding 
canonical ensemble, a full {\it quantitative} agreement for the whole distribution 
is only obtained whenever the level statistics in the bulk of the spectrum is of Wigner-Dyson type, 
corresponding to diffusive eigenstates in quasi-particle space. 

This paper is organized as follows. In Sec.~\ref{sec:model} we define the models 
and the details of the quantum quench protocol. 
In the following two sections we first discuss the spectrum of the Hamiltonian, which provides an insight 
in the localization/delocalization transition (Sec.~\ref{sec:spectral}), and then
we briefly discuss how the long-time asymptotics of different correlation functions is affected 
by the localization/delocalization in many-body space (Sec.~\ref{sec:obs}), thus
summarizing the results of Ref.~\cite{Canovi2011}. 
In Sec.~\ref{sec:pdf} we take a step further: we define the FPDF of the transverse 
and longitudinal spin and discuss their behavior in the asymptotic long-time state attained after
a quench ({\it diagonal ensemble}). Finally, in Sec.~\ref{sec:conc} we draw our conclusions.

\section{Model}\label{sec:model}

Throughout this paper we will consider a quantum quench described by a time-dependent Hamiltonian:
\beq \label{eq:model0}
   \Ham(t) \equiv \Ham_0[g(t)] + \Ham_{ib} \, ,
\eeq
where:
\beq \label{eq:QuenchScheme}
   g(t) = \left\{ \begin{array}{ll} g_0 & {\rm for} \quad t<0    \, ,\\
                                 g   & {\rm for} \quad t \geq 0  \, .   \end{array} \right.
\eeq
The Hamiltonian $\Ham(t)$ is composed of an integrable part $\Ham_0[g(t)]$,  
and an integrability-breaking term given by $\Ham_{ib}$. 
Concerning the integrable part, we will consider the anisotropic spin-1/2 Heisenberg chain 
of length $L$ (also called the XXZ model) with open boundary conditions: 

\beq \label{eq:XXZ}
\Ham_0(J_z) = \sum_{i=1}^{L-1} \left[
   J \left(\sigma^{x}_{i}\sigma^{x}_{i+1}+\sigma^{y}_{i}\sigma^{y}_{i+1}\right)
    + J_{z}\sigma^{z}_{i}\sigma^{z}_{i+1} \right] \,,
\eeq
where $\sigma^{\alpha}_{i}$ ($\alpha=x,y,z$) are the spin-1/2 Pauli matrices on site $i$, 
$J$ is the planar $xy$-coupling, while $J_{z}$ is the nearest-neighbor anisotropy parameter 
in the $z$ direction, which coincides with the parameter $g$ that will be quenched at time $t=0$.
In what follows we take $\hbar = k_{B} = 1$, we adopt $J = 1$ as the energy scale 
and work in the zero total magnetization sector along the $z$ axis.
This model is integrable by Bethe Ansatz~\cite{Nagaosa:book}. 
The zero-temperature phase diagram is characterized by three regions: 
a gapped ferromagnetic phase ($J_{z}<-1$), 
a gapped  antiferromagnetic phase ($J_{z}>1$),  and a gapless critical phase for $-1\leq J_{z}\leq 1$. 
In this gapless region the critical exponents depend on $J_{z}$ and the system 
is characterized by a quasi-long-range order in the $xy$ plane~\cite{Haldane_PRL81}.
In the following we break the integrability of the model by applying a random magnetic field:
\beq\label{eq:IntBreak}
   \Ham_{ib}=  \Delta \sum_{i=1}^{L}h_{i}\sigma^{z}_{i}      \, ,
\eeq
where the quantities $h_i$ are randomly chosen in the interval $[-1,1]$. 

We point out that the model described in Eqs.~\ref{eq:XXZ},\ref{eq:IntBreak} is equivalent 
to a system of hard-core bosons, as one can show by applying the Jordan-Wigner transformation. 
Therefore it is also interesting from an experimental perspective, since cold-atom gases 
in one dimension and at low densities behave like impenetrable bosons~\cite{Bloch_RMP08}. 

After investigating the transition from integrability to non-integrability, we will address 
the long-time behavior of the system following a sudden quench of $J_z$. 
In this context, the knowledge of a substantial part of the spectrum is required, 
making it necessary to resort to a standard exact diagonalization technique.
We will diagonalize Hamiltonian systems with up to 14 sites, and only consider 
the zero magnetization sector, thus working in Hilbert spaces with up to 3432 states.

The zero-temperature phase-diagram of the XXZ model in presence of disorder 
is well established~\cite{Doty_PRB92}.
Much less is known at infinite temperature: the phase-diagram has been conjectured to be composed of two phases, 
a non-ergodic many-body localized phase (in real space) at $\Delta > \Delta^{\rm crit}$, 
and an ergodic one at $\Delta < \Delta^{\rm crit}$, where $\Delta^{\rm crit} \sim 6\div 8$ 
at $J_z=1$~\cite{Pal_PRB10, Znidaric_PRB08}. 
Our results indicate the presence of a second non-ergodic localized phase (in quasi-particle space) 
for $\Delta \equiv \Delta^\star $ close to zero that crosses over to the ergodic phase upon increasing $\Delta$. 
The fate of this crossover in the thermodynamic limit and the eventual value of the critical $\Delta^\star$ 
are yet to be determined~\footnote{While for the parameters used in this paper the low-lying eigenstates are localized 
in the thermodynamic limit, in the following we consider systems sizes smaller than the localization length.}.  

As the strength of $\Delta$ is varied, the system deviates from integrability, as discussed more 
in detail in Sec.~\ref{sec:spectral}, where the spectrum and the properties of the eigenstates 
are quantitatively investigated. Indeed, as the level statistics changes from Poissonian 
(typical of integrable systems) to Wigner-Dyson (characterized by the level repulsion 
of non-integrable systems), the eigenstates are also modified. 
More specifically, when the system is integrable they are localized in quasi-particle space, 
while they become diffusive as the system becomes non-integrable. This transition affects
the dynamics of the system. As summarized in Sec.~\ref{sec:obs}, there is a connection between 
the onset of thermalization and the many-body localization transition of the eigenstates
(for further details, we refer the reader to Ref.~\cite{Canovi2011}).
Later we will go one step forward and investigate how the localization/delocalization transition 
emerges in the FPDF of operators~\cite{Berges2004, Moeckel_PRL08, Moeckel_NJP10}.

\section{Spectral characterization of the integrability-breaking crossover} \label{sec:spectral}

In this section we follow the approach of  Ref.~\cite{Canovi2011} and consider the properties of the eigenvalues 
and eigenvectors of the Hamiltonian in Eq.~\ref{eq:model0} for a fixed value of the anisotropy $J_{z} = 0.5$, 
and show that the integrability breaking is associated with a localization/delocalization transition 
in quasi-particle space.

\subsection{Statistics of the energy level spacings}

In finite-size systems, it is commonly believed 
that the statistical distribution of the energy spacings of the quantum energy levels
directly reflects the integrability properties of the model~\cite{Caux_JSTAT11}.
In particular an integrable quantum system is typically signaled by the presence of 
a Poissonian statistics in the distribution of its level spacings $\{ s_n \}$, 
$s_n \equiv E_{n+1}-E_n$ being the spacing between two adjacent levels normalized to the average level spacing,
\beq \label{eq:poisson}
P_{\rm P} (s)=e^{-s}\;.
\eeq
Physically this means that the eigenvalues of the Hamiltonian within a given symmetry sector are allowed to cluster. 
On the contrary, for non-integrable systems level crossing is inhibited. 
The level statistics has a Wigner-Dyson (WD) distribution, which embeds the level repulsion 
in a power-law: $\lim_{s\to0} P_{\rm WD} (s)~\sim~s^{\gamma}$.  
More precisely, when anti-unitary symmetry is preserved, 
as in the present case, the statistics is described by a Gaussian Orthogonal Ensemble~\cite{Haake}:
\beq\label{eq:wigner}
P_{\rm WD} (s)=\frac{\pi s}{2}e^{-\frac{\pi s^{2}}{4}}\;.
\eeq
By tuning the parameter $\Delta$, the XXZ chain undergoes a transition from Poissonian to WD statistics. 
In a finite-size system, this transition takes the form of a smooth crossover
which can be studied within a specific energy shell, by means of the following
level spacing indicator (LSI)~\cite{Jacquod_PRL97}:
\beq \label{eq:eta}
\eta_{w} (E) \equiv \frac{\int_{0}^{s_{0}}[ P_{[E,E+W]}(s)-P_{\rm P}(s)] ds}
                 {\int_{0}^{s_{0}}[ P_{\rm WD}(s)-P_{\rm P}(s)] ds} \;,
\eeq
where $P_{[E,E+W]}(s)$ is the level statistics computed in the window $[E,E+W]$,
while $s_0$ is the first intersection point of $P_{\rm P} (s)$ and $P_{\rm WD} (s)$.

\begin{figure}[!b]
  \begin{center}
    \includegraphics[width=0.6\columnwidth]{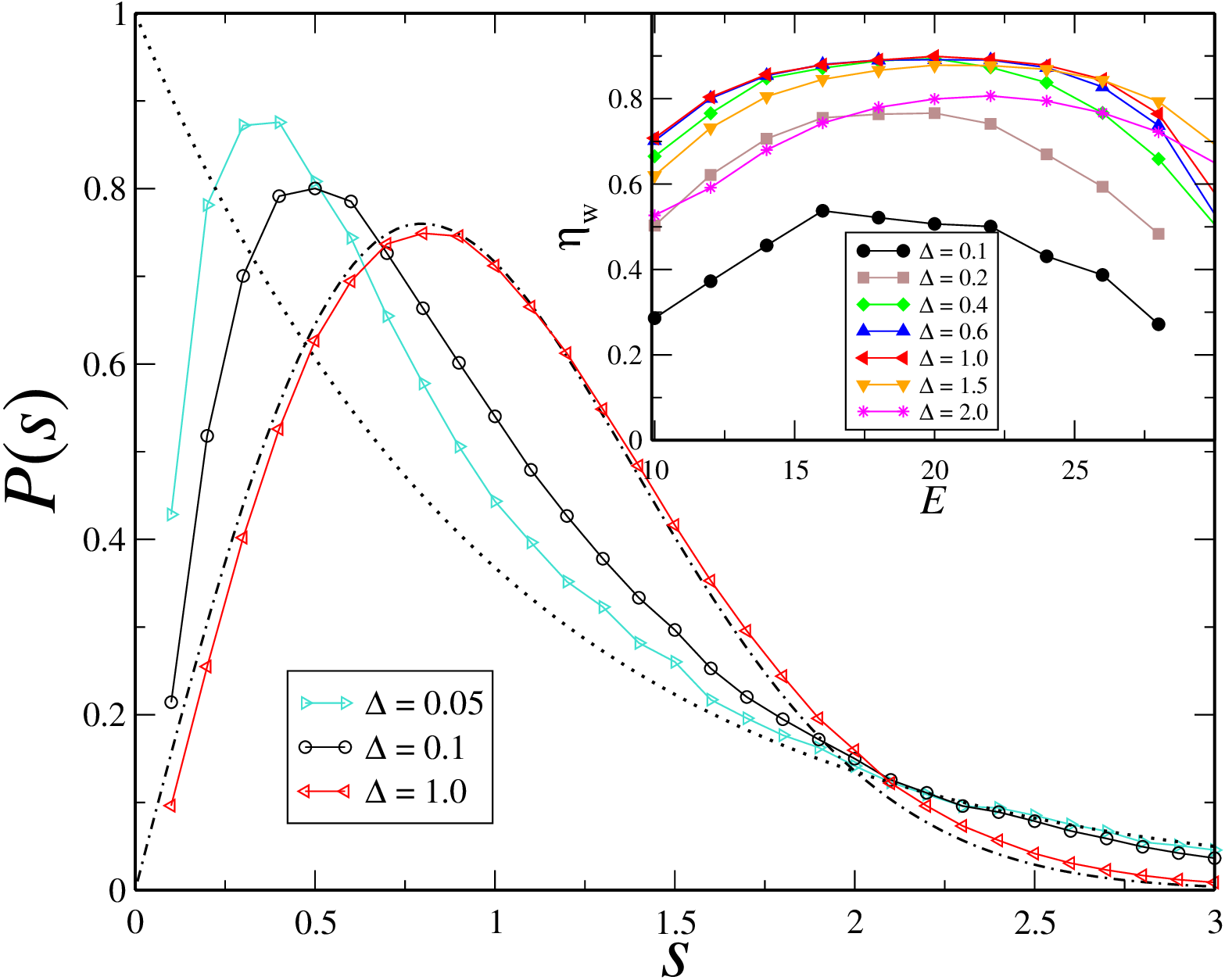}
    \caption{Main panel: level spacing statistics for three values of $\Delta$,
      computed for levels with excitation energy $E$ with respect to the 
      ground state up to the cutoff energy $E_{c}=20$ in our units. The black dotted and 
      dashed-dotted lines show the Poisson (Eq.~\ref{eq:poisson}) and the Wigner-Dyson (Eq.~\ref{eq:wigner})
      distributions respectively.
      Inset: LSI as defined in Eq.~\ref{eq:eta}, evaluated in a microcanonical shell of width $W = 2$.
      Following standard techniques adopted in quantum chaos~\cite{Haake}, 
      we performed an unfolding of the energy spectrum for each instance.
      For both plots data are for $L = 14$ and averages are performed over $5000$ disorder instances.}
    \label{fig:EtaW}
  \end{center}
\end{figure}

In Fig.~\ref{fig:EtaW} we show the level spacing distribution function $P(s)$ for the 
levels with an excitation energy less than a given cutoff , i.e. $E<E_{c}$, (see caption of Fig.~\ref{fig:EtaW}).
We see that for small $\Delta$ the distribution is closer to an exponential, while for 
$\Delta=1$ it almost coincides with a WD distribution, and it is shifted towards 
larger values of the spacings. In the inset of Fig.~\ref{fig:EtaW} we show $\eta_w$ 
for different intensities of the integrability-breaking perturbation.
As the strength of the integrability-breaking term increases, the LSI approaches     
values close to unity for $\Delta\sim 1$. 
At large values of $\Delta\gg J_{z}$ the system tends to another integrable 
limit~\cite{Distasio_PRL95, Kudo_PRB04}, 
indeed we can see that $\eta_{w}$ decreases again for $\Delta\gtrsim 1$. We note that only the states
in the middle of the band display level repulsion (i.e. values of $\eta_{w}$ closer to unity), 
while states in extreme regions of the spectrum do not.
This comes as a consequence of the two-body interaction of the system, as opposed to
the behavior that is typically observed in full random matrices~\cite{Gomez_PR11}. 

\subsection{Inverse participation ratio vs. number of available states}

Along with the Poisson-to-WD transition of the level spacing statistics, 
the eigenvectors also undergo a transition in their statistical properties. 
Indeed, when the level statistics is Poissonian and $\Delta$ is small, the eigenstates 
of the Hamiltonian are close to those of the integrable XXZ chain. 
In other words, they are localized in quasi-particle space. 
On the contrary, when the statistics is WD, the eigenstates are delocalized in quasi-particle space. 
This picture can be made quantitative by using the 
{\it inverse participation ratio} 
(IPR)~\cite{Brown_PRE08, Dukesz_NJP09, Santos_PRE10, Santos_PRE10b, Yurovsky2011, Santos_PRL12}. 
Given a pure state $\ket{\psi}$ and an arbitrary basis $\lbrace\ket{n}\rbrace$ with $N$ elements, 
the IPR is defined by~\cite{Haake}:~\footnote{With 
this definition we fix the typo in Ref.~\cite{Canovi2011} where there was 
a wrong factor $1/N$ in the definition. We remark that the results shown in Ref.~\cite{Canovi2011}
are not affected by that mistake.}
\beq
\xi(\ket{\psi}) = \left(\sum_{n=1}^{N}|\langle n|\psi\rangle|^{4}\right)^{-1} \, .
\eeq
If a state is a superposition of $n_{\rm st}$ basis states, the corresponding contribution 
to $\xi$ is of order $n_{\rm st}$.
Interesting results emerge if one considers two different bases:
(i) the ``site basis'' or ``computational basis'' of the eigenvectors of $\sigma^{z}_{i}$ with the constraint 
of zero total magnetization, and 
(ii) the ``integrable basis'' of the eigenstates of the integrable model under
investigation (the XXZ chain with $J_{z}=0.5$ and $\Delta = 0$). 
Given an eigenstate of $\Ham(J_{z})$, we found that the IPR in the integrable basis 
is of the order of unity if $\Delta\ll J_{z}$, and it grows with increasing $\Delta$.
Eventually, for large values of $\Delta$, the eigenstates of $\Ham(J_{z})$ become diffusive 
superpositions, with random phases and amplitudes, of the eigenstates of the integrable model. 
On the contrary, in the site basis, we found an opposite behavior, since the states 
of this basis approach the eigenstates of the system at $\Delta \gg J_{z}$~\cite{Canovi2011}. 

\begin{figure}[!t]
  \begin{center}
    \includegraphics[width=0.85\columnwidth]{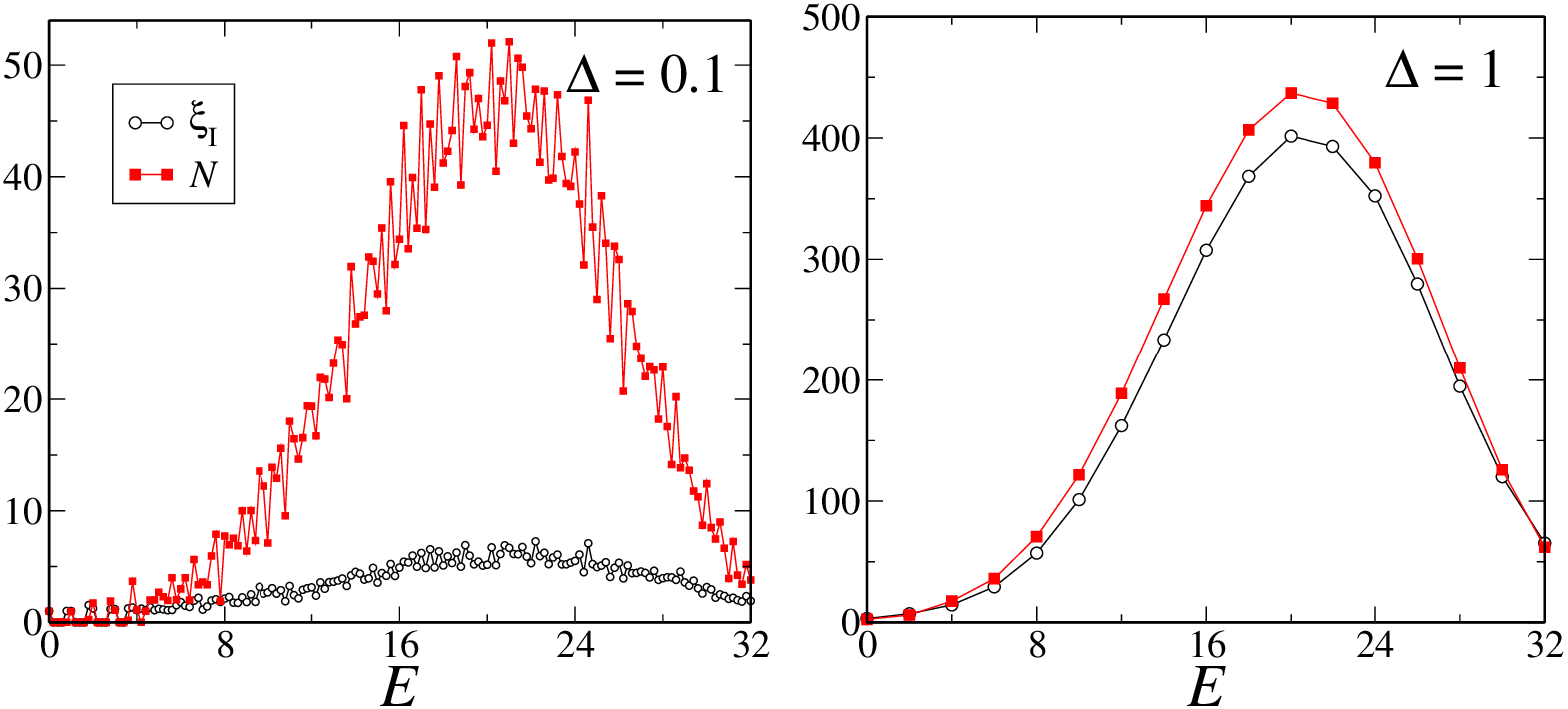}
    \caption{IPR in the integrable basis $\xi_I$ at $\Delta = 0.1$ (left panel), 
      and at $\Delta = 1$ (right panel), as compared to the number of available eigenstates 
      $N$ in an energy window of width $W = 2 \Delta$.
      Data are for a chain of $L=14$ sites. Average over $\sim10^{2}$ disorder realizations.}
    \label{fig:xi_N}
  \end{center}
\end{figure}

We can draw an intuitive picture of the localization/delocalization transition in quasi-particle 
space by comparing the IPR in the integrable basis, in a given microcanonical shell, 
with the number of available eigenstates $N_{[E,E+W]}$ in the same shell. 
This is shown in Fig.~\ref{fig:xi_N}. The microcanonical shell has a width $W=2\Delta\sim V$, 
where $V$ is the typical matrix element of the integrablity-breaking perturbation. 
In quasi-integrable situations ($\Delta \ll 1$, left panel)
the IPR is much lower than the available microcanonical states,
thus meaning that the degree of delocalzation of the system is very low.
On the contrary, in a chaotic situation ($\Delta \sim 1$, right panel), the perturbation
is able to hybridize nearly all the states in the microcanonical energy shell.
We point out that our approach has been also recently adopted in order to identify 
the emergence of chaos in a very similar spin chain model~\cite{Santos_PRL12, Santos_PRE12}.

\section{Long-time dynamics and observables} \label{sec:obs}

\subsection{Energy scales involved in the quench}

\begin{figure}[!t]
  \begin{center}
    \includegraphics[width=0.6\columnwidth]{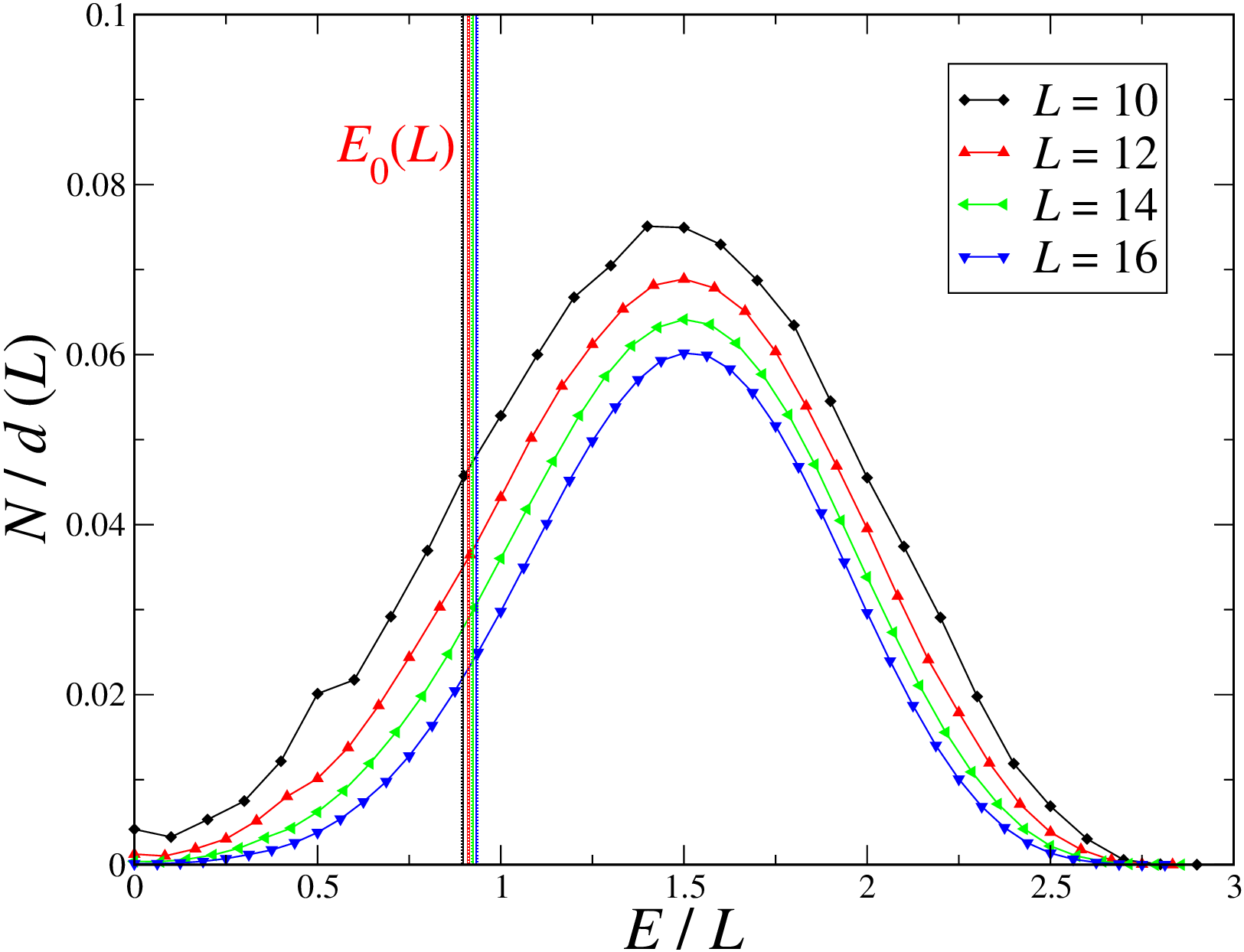}
    \caption{Number of states in each microcanonical shell as compared to the initial energy of the system. 
      The energies in the $x$ axis has been rescaled on the number $L$ of sites,
      while the number of states $N(E)$ is rescaled with the total dimension of the Hilbert space 
      in the zero magnetization sector $d(L)$.
      Vertical lines denote the energy $E_{0}(L)$ after the quench from an antiferromagnetic ground state, obtained with
      $J_{z0}=100$,
      to $J_{z}=0.5$ (the values of $L$ in these vertical lines increase from left to right).
      Different colors refer to different system sizes.
      We used $\Delta=1$ and $W = 2\Delta$. 
      Averages have been performed over $\sim 10^{3}$ random instances of the magnetic field.}
    \label{fig:band}
  \end{center}
\end{figure}

Let us now consider a sudden quench of the parameter $g\equiv J_{z}$, from $J_{z0}$ 
at time $t\leq 0$ to $J_{z}$ at time $t>0$, as in Eq.~\ref{eq:QuenchScheme}. 
We assume that the initial state $\ket{\psi_{0}}$ of the system is the ground state 
of $\Ham(J_{z0})$. Since after the quench the Hamiltonian is time-independent, 
the energy is conserved and is given by $E_{0}=\sandwich{\psi_{0}}{\Ham(J_{z})}{\psi_{0}}$. 
For large values of $J_{z0}$ the ground state of the Hamiltonian is the classical antiferromagnetic 
N\'eel state and the energy density $E_{0}/L$ converges to a constant value, below the middle 
of the spectral band of the final Hamiltonian. 
This means that the eigenstates of $\Ham(J_{z})$ involved in the time evolution 
of the system are only those lying in the first half of the band.
This is shown in Fig.~\ref{fig:band}, where the energy $E_{0}$ after the quench
from an initial antiferromagnetic state (vertical lines) are compared with 
the number of available states in a given microcanonical energy shell.

Following Rossini {\it et al.}~\cite{Rossini_PRL09,Rossini_PRB10}, we then proceed to define 
an effective temperature associated to the quench as the solution of the equality:
\beq\label{eq:teff}
E_{0}\equiv \langle\Ham(J_{z})\rangle_{T_{\rm eff}}={\rm Tr}[\rho(T_{\rm eff})\Ham(J_{z})] \,,
\eeq
where $\rho(T_{\rm eff})$ is the equilibrium density matrix at temperature $T_{\rm eff}$:
\beq
\rho(T_{\rm eff})=\frac{e^{-\Ham(J_{z})/T_{\rm eff}}}{{\rm Tr}[e^{-\Ham(J_{z})/T_{\rm eff}}]} \,.
\label{eq:rhoT}
\eeq
Equation~\ref{eq:teff} can be solved numerically for each realization of disorder and then averaged.
As shown in Ref.~\cite{Canovi2011}, the effective temperature increases 
with the quench strength $\vert J_z - J_{z0} \vert$, and tends to saturate for large values of $J_{z}$, 
since the ground state approaches the antiferromagnetic N\'eel state. 
As an example, for a system of $L=12$ sites and a quench from $J_{z0}=10$ to $J_{z}=0.5$ 
(that are the typical parameters we considered hereafter) the effective temperature 
at $\Delta = 1.0 $ is $T_{\rm eff}=3.4 \pm 0.4$ (units of $J/k_B$), after averaging 
over 200 disorder instances.

\subsection{Asymptotics of observables} \label{Sec:Asympt_obs}

The connection between the localization/delocalization transition and the onset of thermalization 
deeply affects the dynamics following the quantum quench. 
A possible way to test this interplay is to compare the long-time behavior of the system 
with the thermal behavior at the effective temperature $T_{\rm eff}$. 
This can be done at different levels. 
The first possibility is to take a traditional point of view and study the {\it correlation functions} 
of a given observable, as we illustrate now. 
A second more general (to be discussed in the following) approach consists in investigating directly the 
full probability distribution function of selected observables.

\begin{figure}[!t]
  \begin{center}
    \includegraphics[width=0.85\columnwidth]{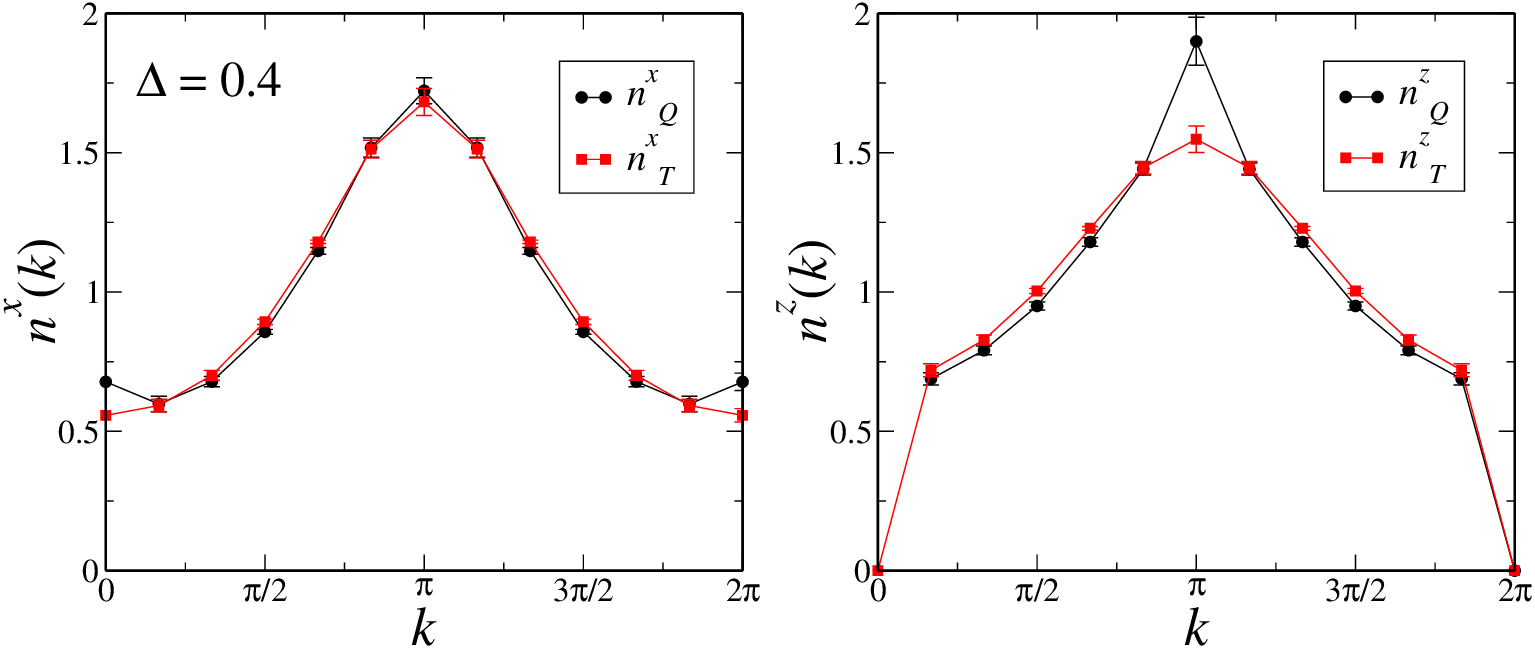}
    \caption{Comparison between the diagonal (black dots) and canonical (red squares) expectation value 
      of the of the two-spin correlation function $n^x(k)$ (left panel) and $n^{z}(k)$ (right panel) 
      as a function of the momentum $k$. 
      Data are for $L=12$ and disorder intensity $\Delta=0.4$.
      Here and in the remaining figures we will always perform a quench from $J_{z0} = 10$ to $J_z = 0.5$.}
    \label{fig:nx}
  \end{center}  
\end{figure}

Let us start by considering the two-spin correlation functions constructed as expectation values of the following
operators:
\beq
n^{\alpha}_{k}\equiv\frac{1}{L}\sum_{j,l=1}^{L}e^{2\pi i(j-l)k/L}\sigma^{\alpha}_{j}\sigma^{\alpha}_{l}, \quad (\alpha=x,z) \,.
\eeq
The average value predicted by the canonical ensemble at temperature $T_{\rm eff}$ is given by:
\beq
n^{\alpha}_{T_{\rm eff}}(k)\equiv \langle n^{\alpha}_{k}\rangle_{T_{\rm eff}}={\rm Tr}[\rho(T_{\rm eff}) \, n^{\alpha}_{k}]\;.
\label{eq:obs_teff}
\eeq
The asymptotic value after the quench is found from the diagonal ensemble~\cite{Rigol_PRL09, Rigol_NAT08, Rigol_PRA10}:
\beq
n^{\alpha}_{Q}(k)\equiv \lim_{t\to\infty}\sandwich{\psi(t)}{n^{\alpha}_{k}}{\psi(t)}=
\sum_{i}|c_{i}|^{2}\sandwich{\phi_{i}}{n^{\alpha}_{k}}{\phi_{i}}\;,
\label{eq:obs_diag}
\eeq
where $\ket{\psi(t)}=e^{-i\Ham(J_{z})t}\ket{\psi_o}$ is the state of the system at time $t$, 
while $c_{i} = \langle \psi_o \vert \phi_i \rangle$ is the $i$-th component of the initial
state $\ket{\psi_{0}}$ in the basis of the eigenstates $\lbrace\ket{\phi_{i}}\rbrace$ 
of the final Hamiltonian $\Ham(J_{z})$.
In Fig.~\ref{fig:nx}, we compare the expectation values in the diagonal ensemble (black data) 
of these operators with those predicted by the canonical ensemble at the temperature $T_{\rm eff}$ (red data). 
We have chosen $\Delta=0.4$, so that the system is still close to integrability
and one can appreciate the discrepancies with the thermal behavior only 
for the correlator of $\sigma_{k}^{z}$ (right panel). On the other hand for $\sigma^{x}_{k}$ (left panel)
a quantitative agreement with the thermal ensemble is observed.
Note that the different behavior of the two observables is most visible at $k=\pi$, 
where the system is less sensitive to boundary effects. 
As discussed in Ref.~\cite{Canovi2011}, the different behaviors of these correlators at long times 
could be related to the different nature of the operators involved in the low energy 
limit~\cite{Rossini_PRL09, Rossini_PRB10}. Indeed, in the low energy limit~\cite{Nagaosa:book}, 
where the XXZ chain critical phase
maps onto a Luttinger liquid and quasi-particles are approximately free bosons, $\sigma_{k}^{z}$ turns out to 
be a local operator, coupling a finite number of quasi-particle states, 
while $\sigma^{x}_{k}$ couples all the states and is thus non-local.  

\begin{figure}[!t]
  \begin{center}
    \includegraphics[width=0.6\columnwidth]{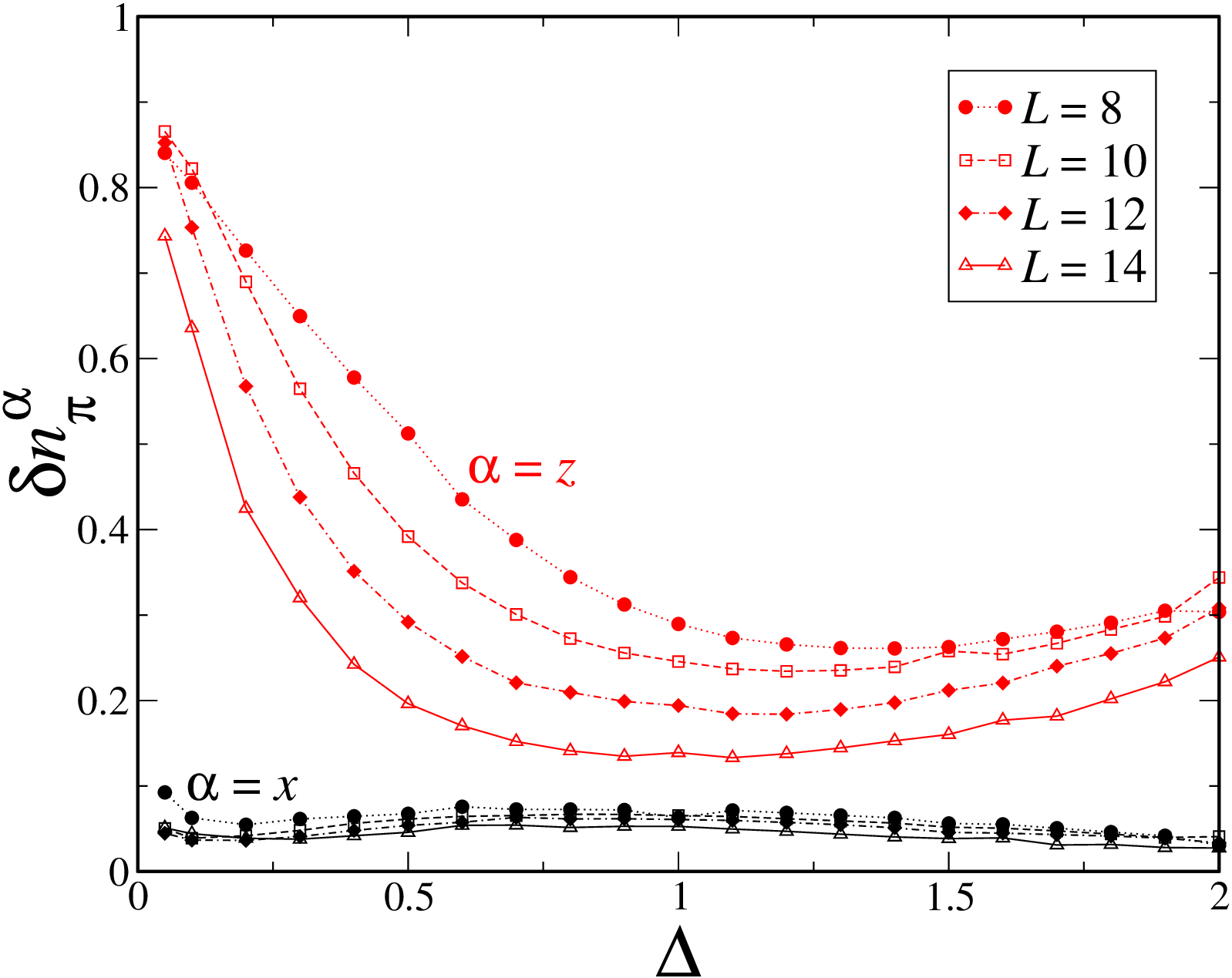}
    \caption{Residuals $\delta n^{x}_\pi$ (black curves) and $\delta n^{z}_\pi$ (red curves) 
      between the diagonal and the canonical ensemble predictions.
      Different symbols refer to different system sizes, as depicted in the caption.
      Averages over 200 instances are performed.}
    \label{fig:deltaN}
  \end{center}  
\end{figure}

To be more quantitative, we define the absolute ``residual'' of the operator  $ n^{\alpha}_{k}$ 
between the diagonal and the canonical ensemble: 
\beq
   \delta n^{\alpha}_{k}=\vert n^{\alpha}_{Q}(k)-n^{\alpha}_{T_{\rm eff}}(k)\vert\;.
   \label{eq:diffObs}
\eeq
In Fig.~\ref{fig:deltaN} we show such quantity at $k=\pi$, as a function of $\Delta$. 
The residual for the non-local operator, $\delta n^{x}_{\pi}$, does not depend substantially 
on the strength of $\Delta$ nor on the size of the system. 
On the contrary, the residual of the local operator, $\delta n^{z}_{\pi}$, decreases significantly 
as the system departs from integrability and shows a minimum for $\bar\Delta =1$. 
For larger values of $\Delta$ the residual $\delta n^{z}_{\pi}$ starts to grow. 
This is consistent with the fact that for large values of $\Delta$ the system approaches 
another integrable limit. Moreover, the quantity $\delta n^{z}_{\pi}$, and in particular 
the position of the minimum $\bar \Delta$ are size-dependent.
Given the numerical limitations of exact diagonalization, we cannot guess what is 
the limit of $\bar\Delta(L)$ when $L\to\infty$. 
The value $\bar \Delta$ of the perturbation strength at which the system is closer 
to a thermal behavior, at a given size, corresponds to the situation in which
delocalization in Fock space is most pronounced.

\section{Full probability distribution functions} \label{sec:pdf}

The study of FPDFs of observables, instead of only their average value, has attracted 
a lot of interest, both theoretically~\cite{Kitagawa2011,Polkovnikov_PNAS06,Gritsev_NP06,Kitagawa_PRL10} 
and, very recently, experimentally~\cite{Gring2011}.
Here we want to study the probability distribution function of the transverse and of the longitudinal 
spin of a subsystem ${\cal S}$ of size $R \leq L$.

The $xy$ transverse-spin magnitude of a subsystem ${\cal S}$ made by $R$ spins is given by:
\begin{eqnarray}
  \nonumber
  (\hat S^{\perp}_{R})^{2} & = & \left\vert \hat S^{x}_{R}+i\hat S^{y}_{R} \right\vert^{2} 
  = \Bigg| \sum_{j \in \cal S}(\hat S^{x}_{j}+i \hat S^{y}_{j}) \Bigg|^2 \\
  & = & \Bigg( \sum_{j \in \cal S}\sigma^{+}_{j} \sigma^{-}_{j} \Bigg) 
      + \Bigg( \sum_{j<l \in \cal S} \left( \sigma^{+}_{j} \sigma^{-}_{l} + \sigma^{-}_{j}\sigma^{+}_{l}\right) \Bigg) 
   \label{eq:transv} 
\end{eqnarray}
where $S^{\alpha}_{i}=\frac{1}{2}\sigma^{\alpha}_{i}$ is a spin operator on site $i$ and direction $\alpha$, 
while spins inside ${\cal S}$ are chosen contiguously in the central portion of the total system
and are identified by the indexes $j = \frac{L}{2} - \frac{R}{2} +1, \ldots , \frac{L}{2} + \frac{R}{2}$.
This is exactly the same quantity considered in Ref.~\cite{Kitagawa2011}, where the full distribution 
of quantum noise in Ramsey interference experiments is studied in detail.

The $z$ component of the spin for a subsystem of size $R$ is given by:
\beq
\sigma^{z}_{R} \equiv \sum_{j \in \cal S} \sigma^{z}_{j} 
\eeq
Since the eigenvectors of $\sigma^{z}_{R}$ coincide with the states of the computational basis, 
the computational cost of evaluating the probability distribution of this quantity 
is less than that for $(\hat S^{\perp}_{R})^{2}$.

\subsection{Asymptotics of the FPDFs}

Similarly to what we have done in Sec.~\ref{Sec:Asympt_obs} for the correlation functions, 
we now consider the behavior of the FPDFs in the diagonal~\cite{Rigol_PRL09,Rigol_NAT08} 
and in the canonical ensemble at the effective temperature $T_{\rm eff}$, and compare the results.

The FPDF can be evaluated in the {\it canonical ensemble} at the effective 
temperature $T_{\rm eff}$ according to
\begin{eqnarray}
   P_{T_{\rm eff}}^{\alpha}(s_{R})& = & \sum_{n} \bigg( {\rm Tr} \left[ \rho(T_{\rm eff}) \, {\cal P}^\alpha_{R,n} \right] \bigg)
   \, \delta(s_{R}-s_{R,n}) \nonumber\\
   & = & \sum_{n} \,p^{\alpha}_{T_{\rm eff}}(s_{R,n})\,\delta(s_{R}-s_{R,n}) \,, \label{eq:pcanon}
\end{eqnarray}
where $p^{\alpha}_{T_{\rm eff}}(s_{R,n}) \equiv {\rm Tr} \left[ \rho(T_{\rm eff}) \, {\cal P}^\alpha_{R,n} \right]$
denotes the thermal expectation value of the projector ${\cal P}^\alpha_{R,n}$ on a given eigenstate 
of $\sigma^\alpha_{R}$ ($\alpha = \perp, z$) corresponding to the eigenvalue $s_{R,n}$
(see the correspondence with Eq.~\ref{eq:obs_teff}).
%
%
On the other side, analogously to what we did in Eq.~\ref{eq:obs_diag},
we define the FPDF in the {\it diagonal} ensemble as
\begin{eqnarray}
   P_{Q}^{\alpha}(s_{R}) & = & \sum_{n} \bigg( \sum_{i} |c_i|^{2} \bra{\phi_i} {\cal P}^\alpha_{R,n} \ket{\phi_i} \bigg) \,
   \delta (s_{R}-s_{R,n}) \nonumber \\
   & = & \sum_{n} \,p^{\alpha}_{Q}(s_{R,n})\,\delta(s_{R}-s_{R,n}) \,. \label{eq:pquench}
\end{eqnarray}

The comparison between the two ensembles can be made quantitative by computing 
the absolute difference (see Eq.\ref{eq:diffObs}):
\beq
   \delta p^{\alpha}(s_{R,n})=|p^{\alpha}_{Q}(s_{R,n})-p^{\alpha}_{T_{\rm eff}}(s_{R,n})|
\eeq
The latter quantity depends on $R$ and on the values of $s_{R}$ specific to each $R$.
We may be interested in comparing the FPDF's at different distances: in this case a useful quantity 
is the integrated difference, which is given by:
\beq
   \label{eq:intdiff}
   \Delta p^{\alpha}(R) = \sum_{s_{R,n}}' \delta p^{\alpha}(s_{R,n})/\nu(R)
\eeq
where $\sum_{s_{R,n}}'$ denotes the sum over the eigenvalues of $\sigma^\alpha_{R}$ 
compatible with the zero total magnetization condition, and $\nu(R)$ is their number.
In the following, we will average all these quantities over disorder, 
and identify their uncertainty with the standard deviation. 
We are going to show our results for a quench from $J_{z0}=10$ to $J_{z}=0.5$
in a system with $L = 12$ spins.

\begin{figure}[!t]
  \begin{center}
    \includegraphics[width=0.8\columnwidth]{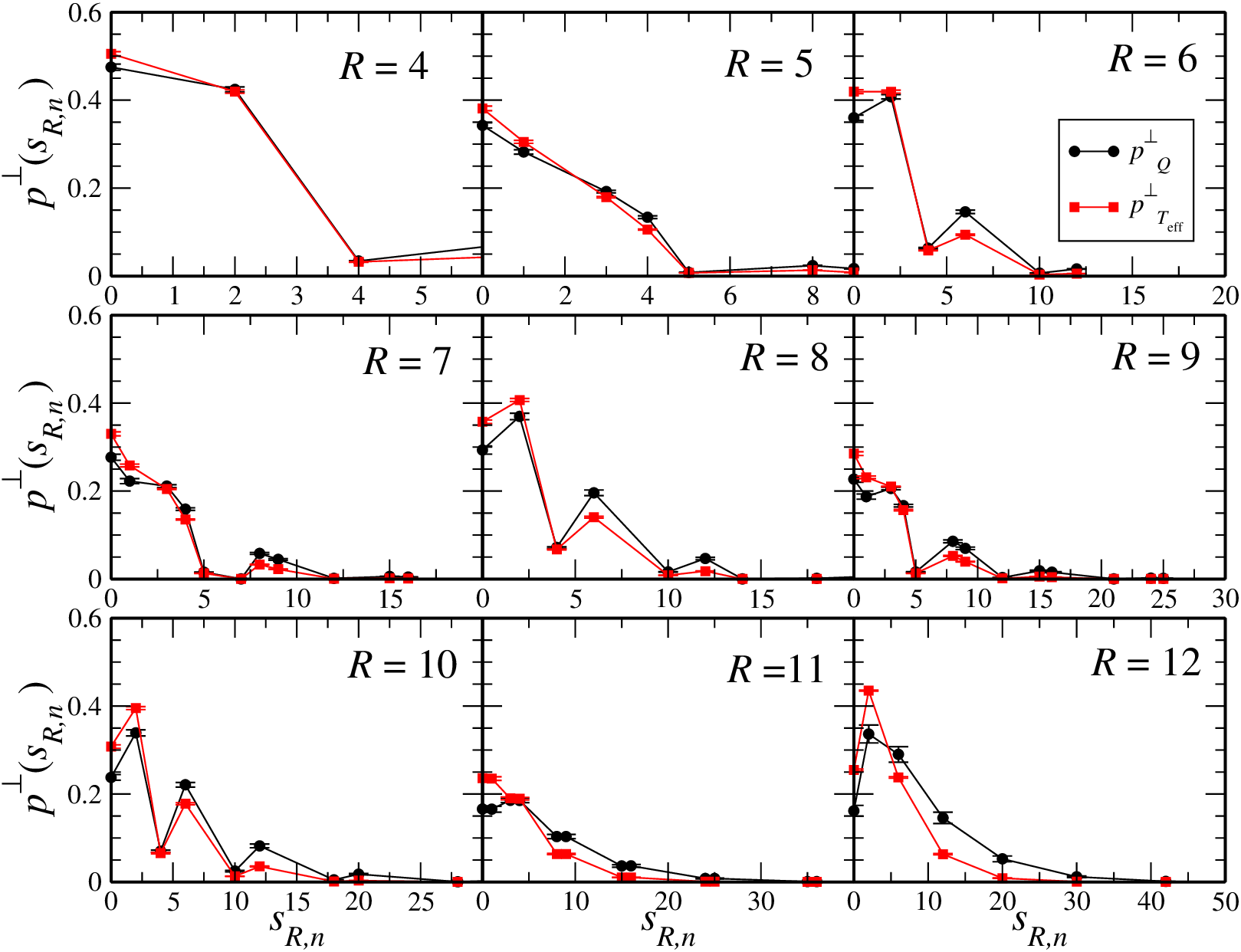}
    \caption{Probability distribution of the transverse spin $p^{\perp}(s_{R,n})$ for a system 
      of $L=12$ sites and $\Delta=0.1$. Different panels refer to different sizes of the subsystem.
      Here and in the remaining figures averages are performed over $500$ disorder realizations.}
    \label{fig:probt0.1}
    \vspace*{0.5cm}
    \includegraphics[width=0.8\columnwidth]{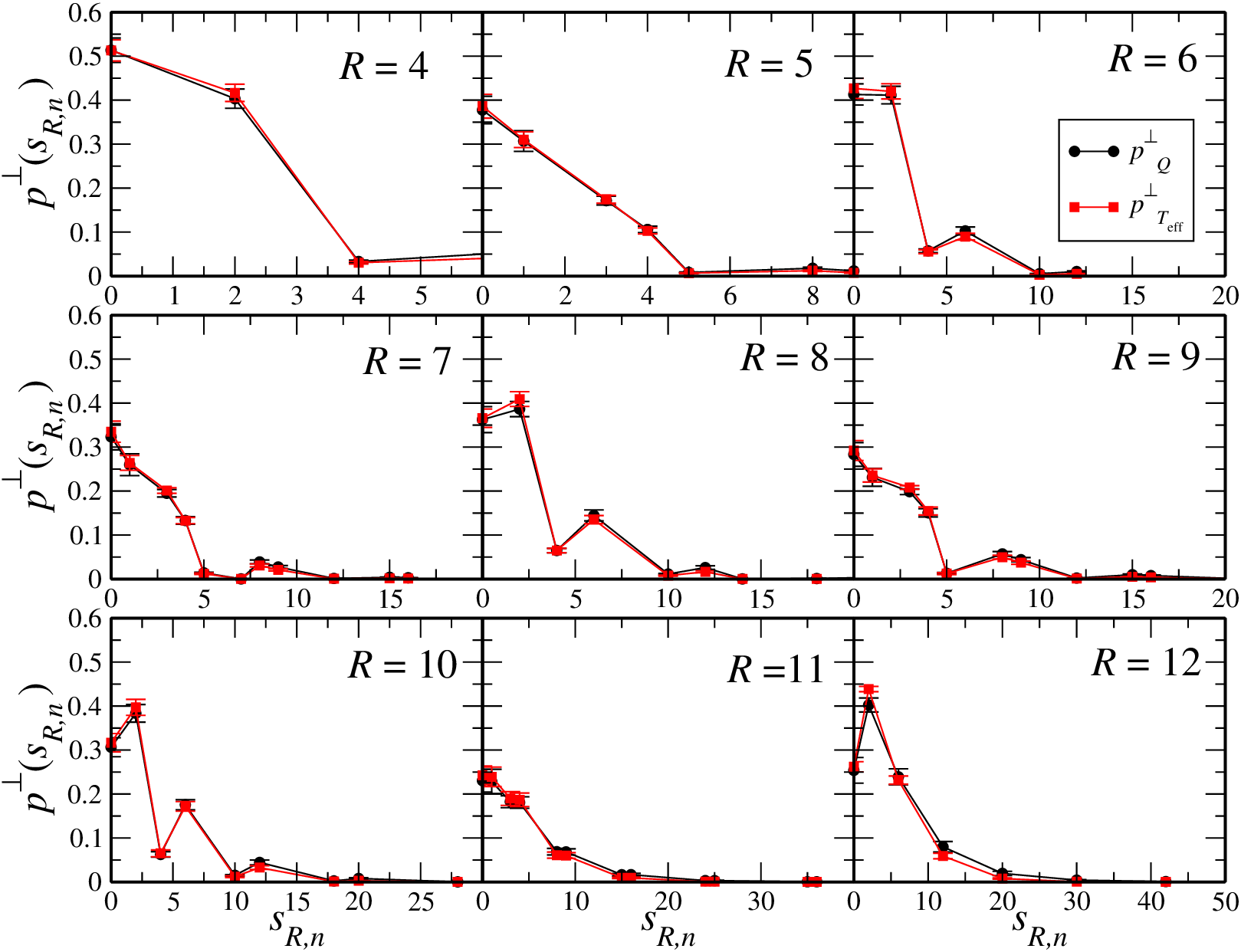}
    \caption{Same as Fig.~\ref{fig:probt0.1}, but for $\Delta=0.5$.}
    \label{fig:probt0.5}
  \end{center}
\end{figure}
\begin{figure}[!t]
  \begin{center}
    \includegraphics[width=0.8\columnwidth]{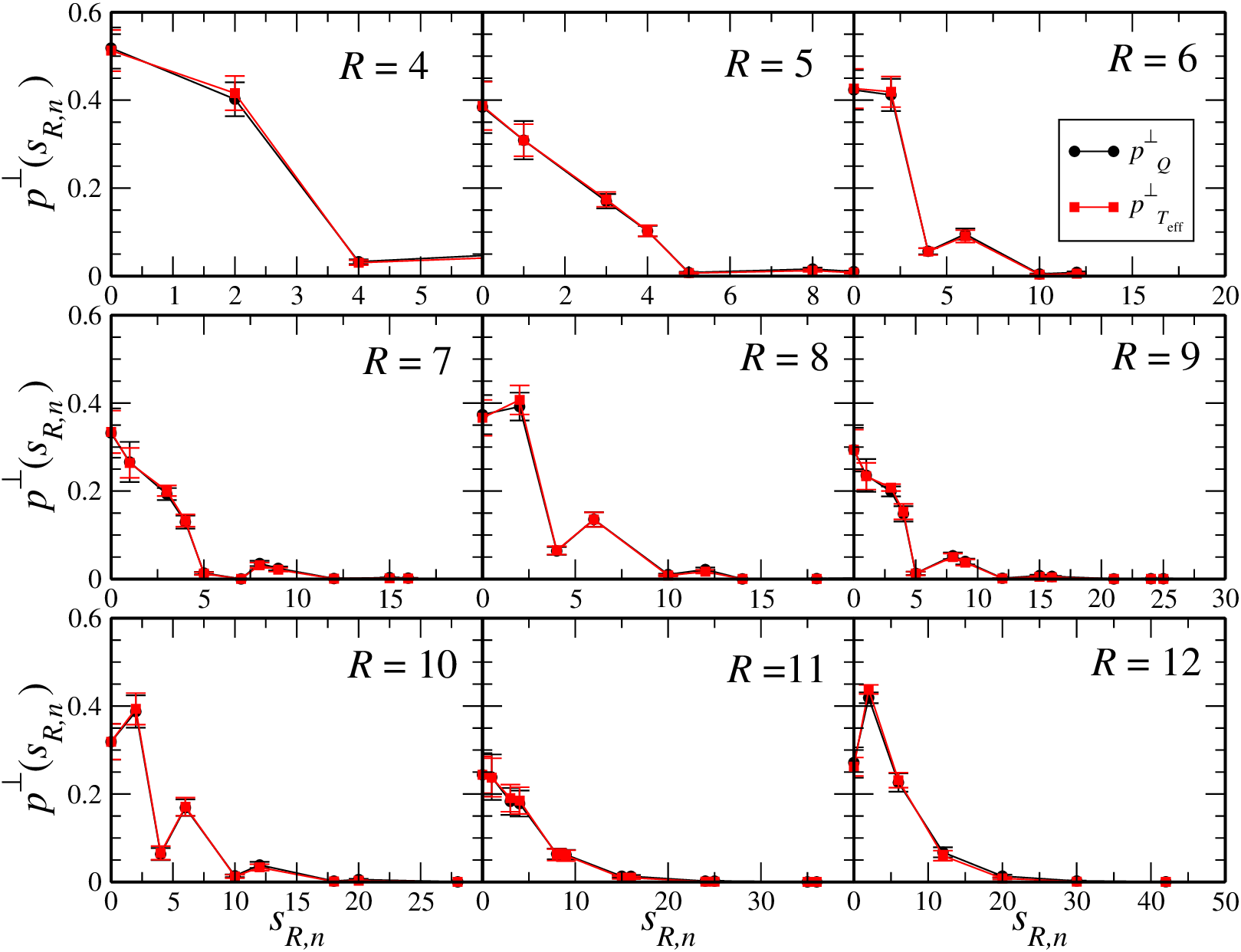}
    \caption{Same as Fig.~\ref{fig:probt0.1}, but for $\Delta=1.0$.}
    \label{fig:probt1.0}
  \end{center}
\end{figure}

\begin{figure}[!t]
  \begin{center}
    \includegraphics[width=0.8\columnwidth]{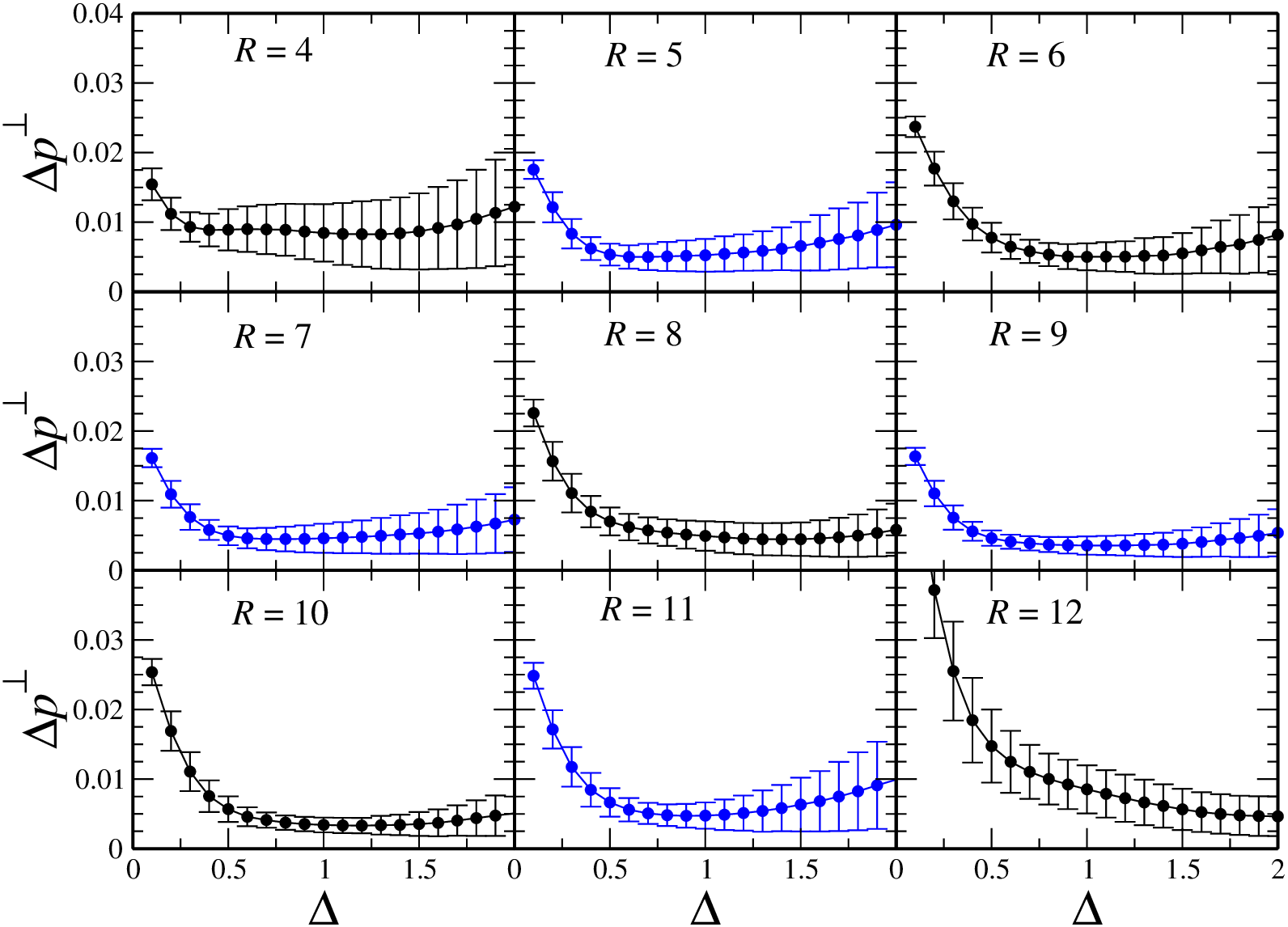}
    \caption{Integrated difference of the transverse spin $\Delta p^{\perp}(R)$ for a system of $L=12$ sites.}
    \label{fig:intdiff12trans}
  \end{center}
\end{figure}

In Fig.~\ref{fig:probt0.1},~\ref{fig:probt0.5} and~\ref{fig:probt1.0} we show the probability
distribution of the transverse spin $(\hat S^{\perp}_{R})^{2}$ at $\Delta= 0.1$, $\Delta= 0.5$ 
and $\Delta=1.0$, respectively. In all the cases, the distribution functions have qualitatively 
the same pattern in both the diagonal and the canonical ensemble.
But when disorder is small, e.g. for $\Delta=0.1$ in Fig.~\ref{fig:probt0.1}, 
there is a visible discrepancy. Increasing the intensity of the disorder a quantitative agreement 
is established, until for $\Delta=1$, that is, where the system is furthest from integrability, 
the two distributions perfectly overlap (Fig.~\ref{fig:probt1.0}). 
%
The discrepancy between the two distributions is quantified by means of the integrated difference 
$\Delta p^{\perp}$ defined in Eq.~\ref{eq:intdiff}, which also allows to compare the distributions for different 
values of $R$. The results are shown in Fig.~\ref{fig:intdiff12trans}. 
The common feature for all values of $R$ is that, when the disorder intensity is small, i.e. $\Delta \ll J_{z}$, the discrepancy
$\Delta p^{\perp}$ is larger and more sensitive to the size of $R$ than for bigger $\Delta$. However it is not clear from our results
 what the dependence on 
$R$ should look like in the thermodynamic limit. Unfortunately with our
simulations we can only work with very small sizes $L$, such that the smallest $R$ is only one order of magnitude less that $L$.
%

\begin{figure}[!t]
  \begin{center}
    \includegraphics[width=0.8\columnwidth]{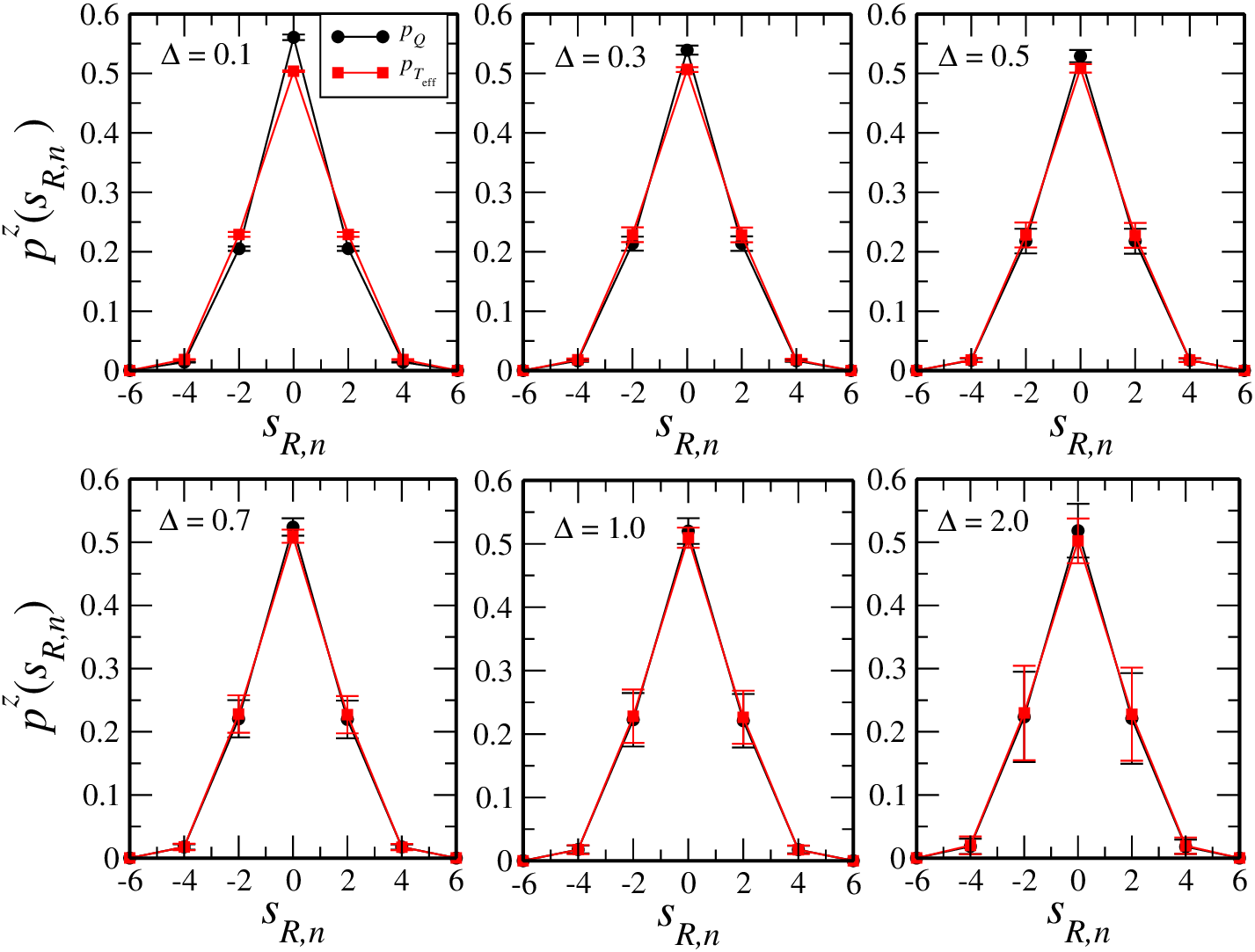}
    \caption{Probability distribution function of $\sigma^{z}_{R}$ in the diagonal (black circles) 
      and canonical (red squares) ensemble. Data for $L=12$ sites, $R=6$ and several values 
      of the disorder intensity.}
    \label{fig:prob12}
    \vspace*{0.5cm}
    \includegraphics[width=0.8\columnwidth]{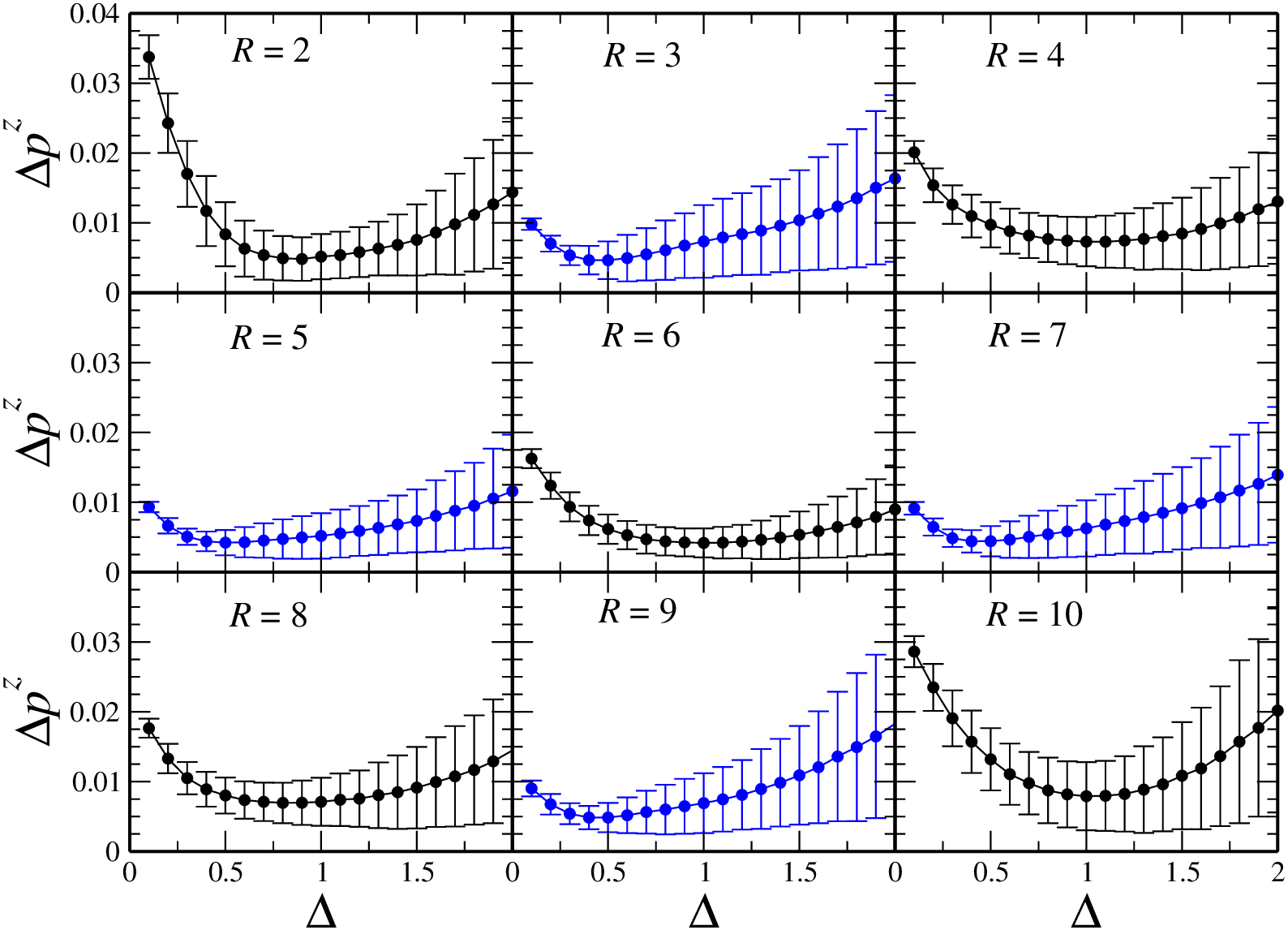}
    \caption{Integrated difference $\Delta p^{z}(R)$ for a system of $L=12$ sites.}
    \label{fig:intdiff12}
  \end{center}
\end{figure}

In Fig.~\ref{fig:prob12} we show the FPDF of the longitudinal spin $\sigma^{z}_{R}$ 
in the diagonal and the canonical ensemble for a system of $L=12$ sites and $R=6$. 
The constraint $\sum_{j}\sigma^{z}_{j}=0$ allows only some eigenvalues of $\sigma^{z}_{R}$, so when the size 
of the subsystem increases from $R=L/2$ to $R=L$  progressively less values of $s_{R}$ are permitted. 
For this reason we choose $R=6$, which is the case with the maximum number of eigenvalues allowed by the symmetries. 
We observe in all the cases that there is a maximum at $s_{R}=0$, which is the most sensitive point 
to the variation of $\Delta$. For odd $R$ (data not shown) there are instead two symmetrical 
eigenvalues, $s_{R}=\pm 1$, where the function has two maxima and which are most suitable 
to check the dependence against $\Delta$. 
From Fig.~\ref{fig:prob12} we see that at small values of $\Delta$ there is a difference 
at $s_{R}=0$ between the two ensembles, while the FPDFs almost coincide for $\Delta \sim 1$. 
The scenario is similar to that of $(S^{\perp}_{R})^{2}$.

In Fig.~\ref{fig:intdiff12} we show the integrated difference $\Delta p^{z}$. 
We see a slight dependence on whether $R$ is even or odd. 
Namely, for even values of $R$ the discrepancy between the ensembles at small values 
of $\Delta$ is more pronounced than for odd $R$. This effect is more visible for small values of $\Delta$. 
For even values of $R$  the difference $\Delta p^{z}$ has its smallest 
value for $\Delta =1$, while for odd values of $R$ the minimum is at $\Delta\sim 0.5$. 
Looking at the error bars, we see that both $p^{z}$ and $\Delta p^{z}$ are generally characterized
by larger fluctuations with respect to the transverse spin, perhaps because 
the spectrum of $\sigma^{z}_{R}$ has a higher degeneracy with respect to $(S^{\perp}_{R})^{2}$. 
However the error bars, both for the transverse and longitudinal spin, are proportional to the disorder intensity.
As we already pointed out for the transverse spin, we cannot fully characterize the dependence of $\Delta p^{z}$ on $R$, because we 
have too small systems. Intuitively, what
we can expect is that the constraint on the total magnetization $\sum_{j}\sigma^{z}_{j}=0$  should relate the situations in which the subsystem 
has $R$ and $L-R$ sites respectively.  On a qualitative level this is also suggested by the panels of Fig.~\ref{fig:intdiff12}.

\section{Conclusions} \label{sec:conc}

In this paper we analyzed the connection between thermalization and many-body localization. 
We studied the long time behavior of a XXZ spin-1/2 chain following a quantum quench, when integrability is 
broken by a random magnetic field.
In particular we addressed the issue of pre-thermalization, not only looking at the behavior of
average values of observables, but also at their FPDF. We found a {\it qualitative agreement} between 
the FPDF in the asymptotic state and that predicted by a thermal distribution
in the whole range of disorder intensity we considered. 
Nevertheless, the situation in which the two FPDFs are {\it quantitatively indistinguishable} 
occurs only when the system is furthest from integrability, that is when the eigenstates 
are diffusive superpositions in quasi-particle space.

\section*{Acknowledgements}

E.C.\ thanks J. Carmelo and C. Evoli for useful discussions and acknowledges financial support from DPG
through project SFB/TRR21.
D.R.\ and R.F.\ acknowledge financial support from EU through SOLID and NANOCTM.
G.E.S.\ acknowledges support by the Italian CNR, through ESF Eurocore/FANAS/AFRI,
by the Italian Ministry of University and Research, through PRIN/COFIN 20087NX9Y7,
by the SNSF, through SINERGIA Project CRSII2 136287\ 1, and by the EU-Japan Project LEMSUPER. 

\section*{References}

\bibliographystyle{iopart-num}
\bibliography{refs}

\end{document}